\documentclass[
 reprint,amsmath,amssymb,aps,
prl,
]{revtex4-2}

\usepackage{graphicx}
\usepackage{dcolumn}
\usepackage{bm}
\usepackage{hyperref}
\hypersetup{colorlinks, linkcolor={blue}, citecolor={blue}, urlcolor={blue}}

\begin{document}
	
\title{Exchange Surface Spin Waves in Type-A van der Waals Antiferromagnets}
\author{Zhoujian Sun$^{1,2}$}
\author{Fuxiang Li$^2$}
\author{Gerrit E. W. Bauer$^{1,3,4}$}
\author{Ping Tang$^1$}
\email{tang.ping.a2@tohoku.ac.jp}	
\affiliation{$^1$WPI-AIMR, Tohoku University, 2-1-1 Katahira, Sendai 980-8577, Japan}
\affiliation{$^2$School of Physics and Electronics, Hunan University, Changsha 410082, China}
\affiliation{$^3$Institute for Materials Research and CSIS, Tohoku University,
2-1-1 Katahira, Sendai 980-8577, Japan} 
\affiliation{$^4$Kavli Institute for Theoretical Sciences, University of the Chinese Academy of Sciences,
Beijing 10090, China}
\date{\today}

\begin{abstract} 
Surface spin waves in the short-wavelength regime enable ultrafast, nanoscale magnon-based devices. Here we report the emergence of surface spin-wave excitations within the bulk magnon band gap of type-A van der Waals antiferromagnets composed of antiferromagnetically coupled ferromagnetic monolayers. In contrast to the magnetostatic Damon-Eshbach modes in magnetic slabs, these surface waves are pure exchange modes owing to the reduced interlayer exchange coupling at surface layers, and thus persist in ultrathin multilayer stacks and at large wave numbers. We show that they can be efficiently excited by electromagnetic waves, with absorption power comparable to or even exceeding that of bulk modes. Moreover, their magnetic stray fields exhibit pronounced even-odd oscillations with the number of monolayers that should be observable by NV-center magnetometry. 
\end{abstract} 

\maketitle
\textit{Introduction.---}The recent discovery of two-dimensional van der Waals (vdW) magnets triggered tremendous research interest due to their potential for highly integrated spintronic devices ~\cite{lee2016ising, gong2017,huang2017,fei2018two,burch2018,bonilla2018strong,li2019intrinsic,li2019intrinsic,zhang2019,khan2020recent,wang2020prospects,yang2021van,wang2022}. They can form heterostructures with other layered materials that display a variety of intriguing phenomena, such as giant tunnel magnetoresistance \cite{song2018,wang2018very,yang2021spin}, magnetoelectric coupling~\cite{arima2011spin,vopson2017measurement,jiang2018electric}, and spin filtering~\cite{kamalakar2016inversion,piquemal2020,zheng2022spin}. Spin waves or their quanta, magnons, are the collective spin excitations that transport spin angular momentum over long distances without Joule heating~\cite{kajiwara2010transmission,fukuhara2013,chumak2014magnon,chumak2015,Wangl2018,Wangl2021}. Of particular interest are surface spin waves with amplitude localized near the surfaces of magnetic materials. In an in-plane magnetized ferromagnetic slab, Damon-Eshbach (DE) surface spin wave~\cite{eshbach1960surface,DAMON1961} emerge when the wave vector is perpendicular to the equilibrium magnetization and propagate unidirectionally, leading to nonreciprocal spin and heat transport \cite{an2013unidirectional,chen2021unidirectional,yu2023chirality}. In a uniaxial antiferromagnet with easy-plane anisotropy, doubly degenerate chiral surface spin waves are connected by time-reversal symmetry \cite{PhysRevB.28.1475,PhysRevLett.45.283}. The DE waves arise from the long-range magnetodipolar interaction in the limit of long wavelengths and thick films, severely limiting their applicability in future miniaturized spin-wave devices. While topological magnon edge states offer an alternative, they require nontrivial magnon band structures with nonzero Berry curvature and typically exist at frequencies much higher than the fundamental magnon gap at the center of the Brillouin zone \cite{zhang2013topological,mook2014edge,owerre2016first,PhysRevLett.123.080501,diaz2019topological,mcclarty2022topological,zhang2024direct}. 


\begin{figure}
	\includegraphics[width=8.6 cm]{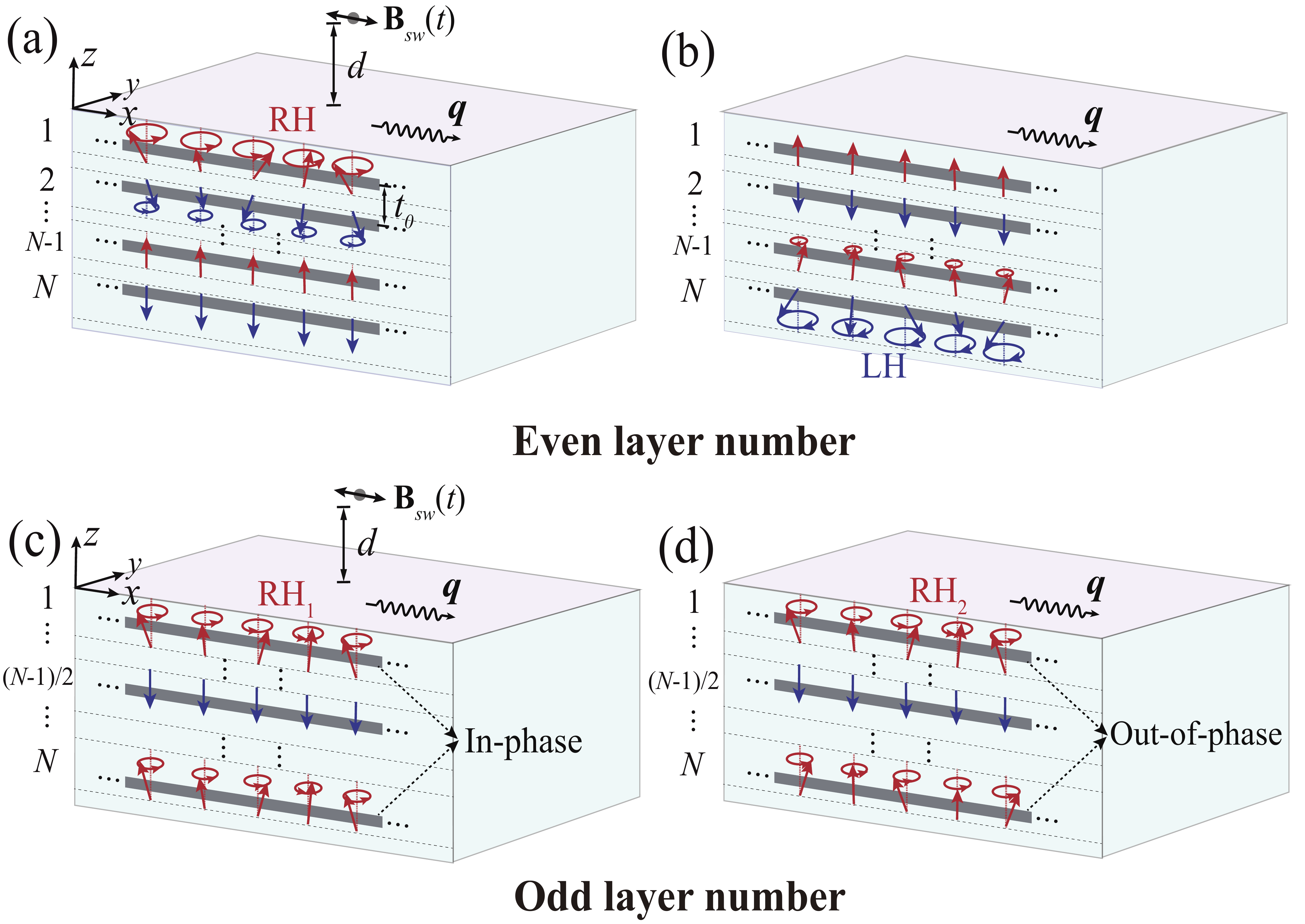}
	\caption{Surface spin waves with wave vector \(\mathbf{q}\) in perpendicularly magnetized A-type van der Waals antiferromagnets. (a), (b) For an even number of monolayers, two surface modes are localized at opposite surfaces with opposite chiralities. (c),(d) For an odd number of monolayers, there are two surface modes of the same (right-handed) chirality with in-phase (c) and out-of-phase (d) spin dynamics between the top and bottom surface amplitudes. The equilibrium magnetization of the upmost ($1$st) monolayer is set to be along the $z$ direction, and RH and LH indicate right- and left-handed precession, respectively.} 
	\label{Fig0}	
\end{figure}

\begin{figure*}
	\includegraphics[width=17cm]{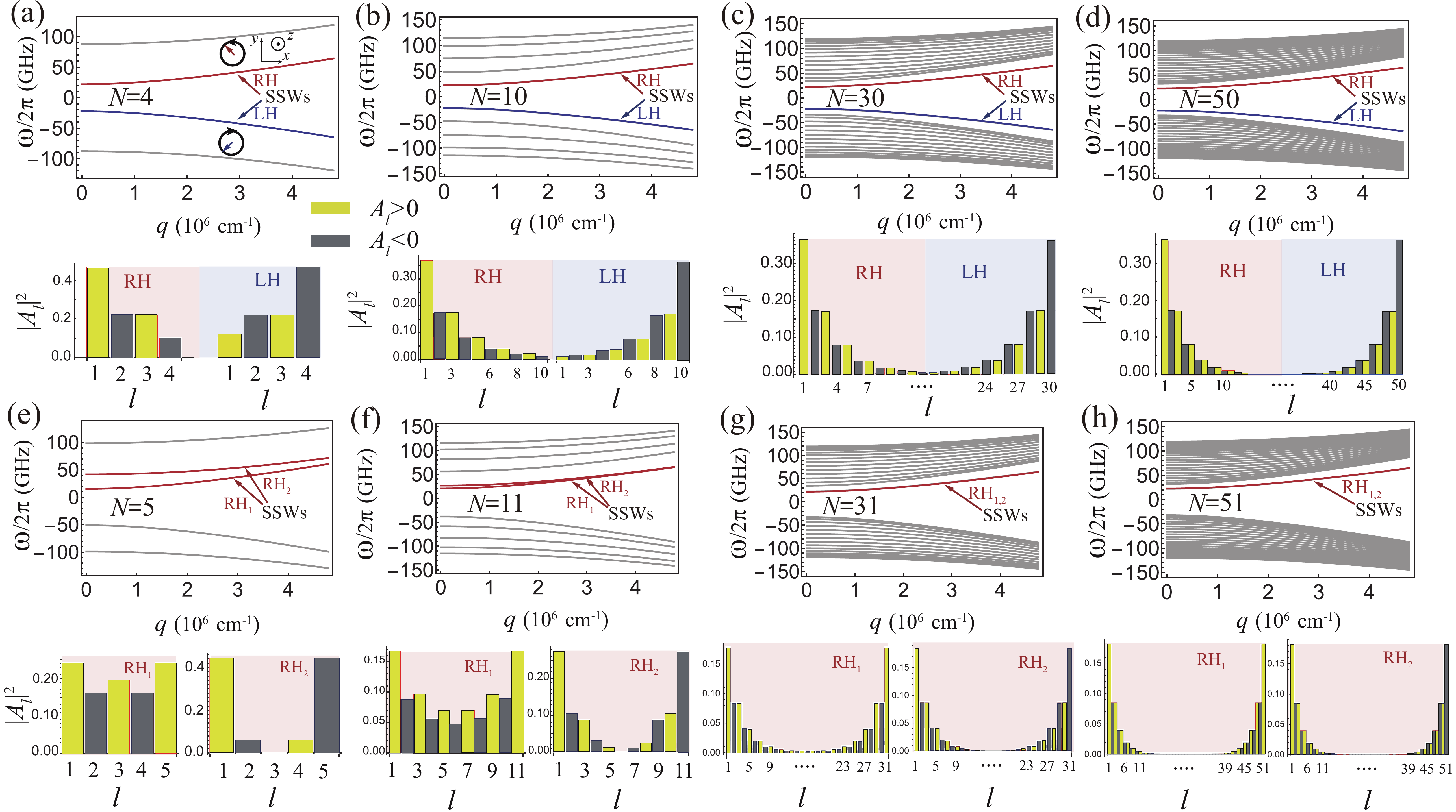}
	\caption{Calculated spin-wave dispersions of van der Waals antiferromagnets CrPS$_4$ with \(N\) monolayers. The lower panels of (a)-(h) show the layer-resolved normalized squared amplitudes \(\vert A_{l}\vert^2\) of two eigenmodes that split off from the bulk-like quasi-continuum with increasing $N$ and correspond in real space to the two surface spin waves (SSWs) shown in Fig.~\ref{Fig0}. The amplitudes \(A_l\) are real-valued with yellow (gray) bars indicating $A_{l}>0$ ($A_{l}<0$). In odd-numbered multilayers, the in-phase and out-of-phase SSWs are associated with symmetric and antisymmetric amplitude profiles with respect to the central layer, respectively. Here, positive and negative frequencies specify right-handed and left-handed modes, respectively, while the physical frequency is \(| \omega |\).}
	\label{Fig1}	
\end{figure*}

In this Letter, we predict a class of novel surface spin wave modes in vdW multilayers with ferromagnetic intralayer and antiferromagnetic interlayer exchange coupling. They emerge from the broken translational symmetry of interlayer exchange interactions at the sample boundaries. They are ``spin-surface" locked, with chiralities determined by the equilibrium magnetization of the surface monolayers, and do not depend on the angle between wave vector and magnetization. Being driven by exchange interactions rather than dipolar forces, they remain robust in the short-wavelength regime and persist even when the multilayer thickness is reduced to a few tens of monolayers. Moreover, we discover an even-odd effect \cite{lounis2008magnetism,machens2013even,zhou2020even,groszkowski2022reservoir,subedi2024even} as a function of the layer number ($N$): When \textit{N} is even, two surface modes are localized at opposite surfaces that are degenerate and propagate with opposite chiralities; when \textit{N} is odd, the amplitudes at both surfaces are coupled through the bulk, leading to the formation of bonding and anti-bonding states that share the same chirality and correspond to in-phase and out-of-phase dynamics of the spins at the top and bottom surfaces, respectively, as illustrated in Fig.~\ref{Fig0}. For perpendicular magnetized CrPS$_4$ with $N\gtrsim30$, these surface modes appear at approximately $2\pi\times22.5\,$GHz, well below the resonance frequencies of the bulk modes \cite{PhysRevB.107.L180403,freeman2025tunable}. They exhibit significant microwave absorption that, while depending on $N$, can even surpass that of the bulk modes. Moreover, their stray magnetic fields oscillate strongly with the layer number, which can be detected, e.g., by NV-center magnetometry~\cite{van2015nanometre,bertelli2020magnetic}.

We focus on vdW magnetic films of $N$ monolayers with out-of-plane equilibrium magnetizations such as CrPS$_4$~\cite{peng2020magnetic}, in which the intralayer and interlayer exchange couplings are ferromagnetic and antiferromagnetic, respectively. When driven by an external \textit{rf} field $\mathbf{h}_{\text{rf}}$, the magnetization dynamics of the $l$th layer is governed by the Landau-Lifshitz-Gilbert (LLG) equation~\cite{landau1992theory,gilbert2004}
\begin{align}
\frac{\partial\mathbf{M}_l}{\partial t}=\gamma \mathbf{H}_{\text{eff}}^{(l)}\times\mathbf M_l+\frac{\alpha_G}{M_s}\mathbf M_l\times\frac{\partial\mathbf M_l}{\partial t}+\gamma \mathbf{h}_{\text{rf}}\times\mathbf{M}_l
\label{LLG}
\end{align}
where $\mathbf{M}_l(\mathbf{r}_{l},t)$ is the magnetization of the $l$th layer, $\vert\mathbf{M}_{l}\vert=M_{s}$, $\gamma$ the absolute value of the gyromagnetic ratio, $\alpha_G$ the Gilbert damping constant, and 
\begin{align}
\mathbf{H}_{\text{eff}}^{(l)}=&\frac{D}{\gamma M_s} \nabla^2 
    \mathbf{M}_l-\frac{H_{E}}{M_{s}}\sum_{l^{\prime}}^{\mathrm{n.n.}}\mathbf{M}_{l^{\prime}}+\frac{H_A}{M_s}(\mathbf{M}_l\cdot \hat{\mathbf{z}})\hat{\mathbf{z}}
    \label{B_eff}
\end{align}
the effective magnetic field in the $l$th monolayer including intralayer ($\sim \nabla^2 
    \mathbf{M}_l$) and nearest-neighbor interlayer ($H_E$) exchange fields, as well as an out-of-plane anisotropy field ($H_{A}$). $D$ is the intralayer spin wave exchange stiffness and  \(l^{\prime}\in\{l+1,l-1\}\) labels nearest-neighbors of layer \textit{l}. For the perpendicularly magnetized magnet of interest, we disregard in Eq.~\ref{B_eff} the long-range dipolar interaction and its effect on the
magnetization dynamics can be well incorporated by an magnetodipolar contribution to $H_{A}$~\cite{PhysRevB.43.6015}.  

Consider a uniform antiferromagnetic ground state magnetization $\mathbf{M}_l=s_{l}M_s\hat{\mathbf{z}}$, where $s_{l}=(-1)^{l+1}$ specifies the equilibrium magnetization direction of the first layer along the positive $z$ direction. A spin wave excitation modulates the magnetization of each layer to $\mathbf{M}_l(\mathbf{r}_l,t)=M_{s}[s_{l}\hat{\mathbf{z}}+\mathbf{m}_{l}(\mathbf{r}_{l}, t)]$, where $\mathbf{m}_l(\mathbf{r}_l,t) \perp \hat{\mathbf{z}}$ is taken to be small. Substituting $\psi_{l}=m_{l}^{x}-im_{l}^{y}$ into Eq.~(\ref{LLG}) leads to the linearized LLG equation
\begin{align}
s_{l}\frac{\partial\psi_{l}}{\partial t}=&i(D\nabla^2-\gamma H_{A}-z_{l}\gamma H_{E})\psi_{l}-i\alpha_{G} \frac{\partial \psi_{l}}{\partial t}\nonumber\\
&-i\gamma H_{E}\sum_{l^{\prime}}^{\mathrm{n.n.}}\psi_{l^{\prime}}+i\gamma (h_{\text{rf}}^{x}-ih_{\text{rf}}^{y}) \label{sLLG}
\end{align}
where $z_{l}$ is the number of neighbors of the $l$th layer, i.e., $z_{l}=1$ ($z_{l}=2$) for surface (bulk) layers. The spin wave (magnon) spectrum and amplitude distribution over the stack follow from Eq.~(\ref{sLLG}) in the absence of damping and driving. Substituting spin-wave solutions, $\psi_{l}(\mathbf{q},t)=A_{l}\exp(i\mathbf{q}\cdot\boldsymbol{\rho}-i\omega t)$, where $A_{l}$ is the layer-dependent amplitude of spin waves with wave vector $\mathbf{q}$, while $\boldsymbol{\rho}$ is the in-plane coordinate. The characteristic matrix equation for spin waves then reads $\hat{\eta}\cdot\hat{H}_{\mathbf{q}}\cdot\hat{A}=\omega \hat{A}$, where $\hat{A}=(A_{1}, A_{2},\cdots, A_{N})^{T}$, $\hat{\eta}=\text{diag}(1,-1, 1,\cdots,)_{N\times N}$ and
 \begin{equation}
\hat{H
}_{\mathbf{q}}=\left(\begin{matrix}
\omega_{\mathbf{q}}+\gamma H_{E} &\gamma H_{E} & 0 &\cdots\\
\gamma H_{E} &\omega_{\mathbf{q}}+2\gamma H_{E} & \gamma H_{E}&0 \\
0 &\gamma H_{E} &\omega_{\mathbf{q}}+2\gamma H_{E}& \gamma H_{E} \\
\cdots&0& \gamma H_{E} & \cdots
\end{matrix}\right)_{N\times N}. \label{H}
\end{equation}   
Here $\omega_{\mathbf{q}}=D\mathbf{q}^{2}+\gamma H_{A}$ is the spin-wave dispersion of an isolated monolayer. The eigenvalues and eigenvectors of the pseudo-Hamiltonian matrix $\hat{\eta}\cdot\hat{H}_{\mathbf{q}}$ are real-valued and encode the spin-wave frequencies and layer-resolved amplitudes, respectively. The interlayer coupling differentiates between interface and bulk layers in the diagonal elements of Eq.~(\ref{H}) and couples neighboring layers via the non-diagonal elements. The former is essential for the emergence of the surface waves, while the latter couples their wave amplitudes at opposite surfaces in odd-numbered layers, as shown below. 

Fig.~\ref{Fig1} presents the spin-wave frequency dispersions for CrPS$_4$ multilayers with $D=6.8 \times 10^{-7}\,$$\text{Hz}\cdot$m$^2$, $H_{A}=0.15$ $\text{T}$, $H_{E}=2.07$ $\text{T}$, and $M_{s}=2.31\times 10^5\,$A/m \cite{PhysRevB.102.024408}. 
From the form of spin-wave solutions, $\mathbf{m}_{l}=A_{l} [\cos(\omega t-\mathbf{q}\cdot\boldsymbol{\rho})\hat{\mathbf{x}}+\sin(\omega t-\mathbf{q}\cdot\boldsymbol{\rho})\hat{\mathbf{y}}]$, positive frequencies indicate right-handed (RH) and negative ones left-handed (LH) modes. For an even number of layers, positive and negative frequency solutions are symmetric and degenerate. For odd numbers, there are always one more positive (RH) than negative (LH) frequency modes, reflecting an uncompensated net magnetization. Remarkably, when $N\gtrsim 30$, two modes split off from the bulk-like quasi-continuum to lower frequencies that in real space are localized at surfaces, as shown in the lower panel of Fig.~\ref{Fig1}. For an even number of layers, the two surface modes are degenerate in frequency and localized at opposite surfaces with \textit{opposite} chiralities. In contrast, for an odd number of layers, the two surface modes are not degenerate with the \emph{same} (RH) chirality, and exhibit symmetric and antisymmetric amplitude profiles relative to the central layer, corresponding to in-phase and out-of-phase spin dynamics between the top and bottom surfaces, respectively [see Fig.~\ref{Fig0}]. When $N$ is small, the symmetric (in-phase or bonding) modes have lower frequency than the antisymmetric (out-of-phase or anti-bonding) ones; the frequency splitting decreases with increasing $N$ and vanishes in the large-$N$ limit. Being independent of the direction of $\mathbf{q}$ and robust at short wavelengths, these states differ fundamentally from the magnetostatic surface waves~\cite{eshbach1960surface,DAMON1961,PhysRevB.28.1475,PhysRevLett.45.283,PhysRevB.27.2955,PhysRevB.31.4465}, but bear analogy with the polarization surface waves in a ferroelectric film with reduced surface pseudo-spin interactions~\cite{cottam1984theory}.

\begin{figure}
	\includegraphics[width=8.6 cm]{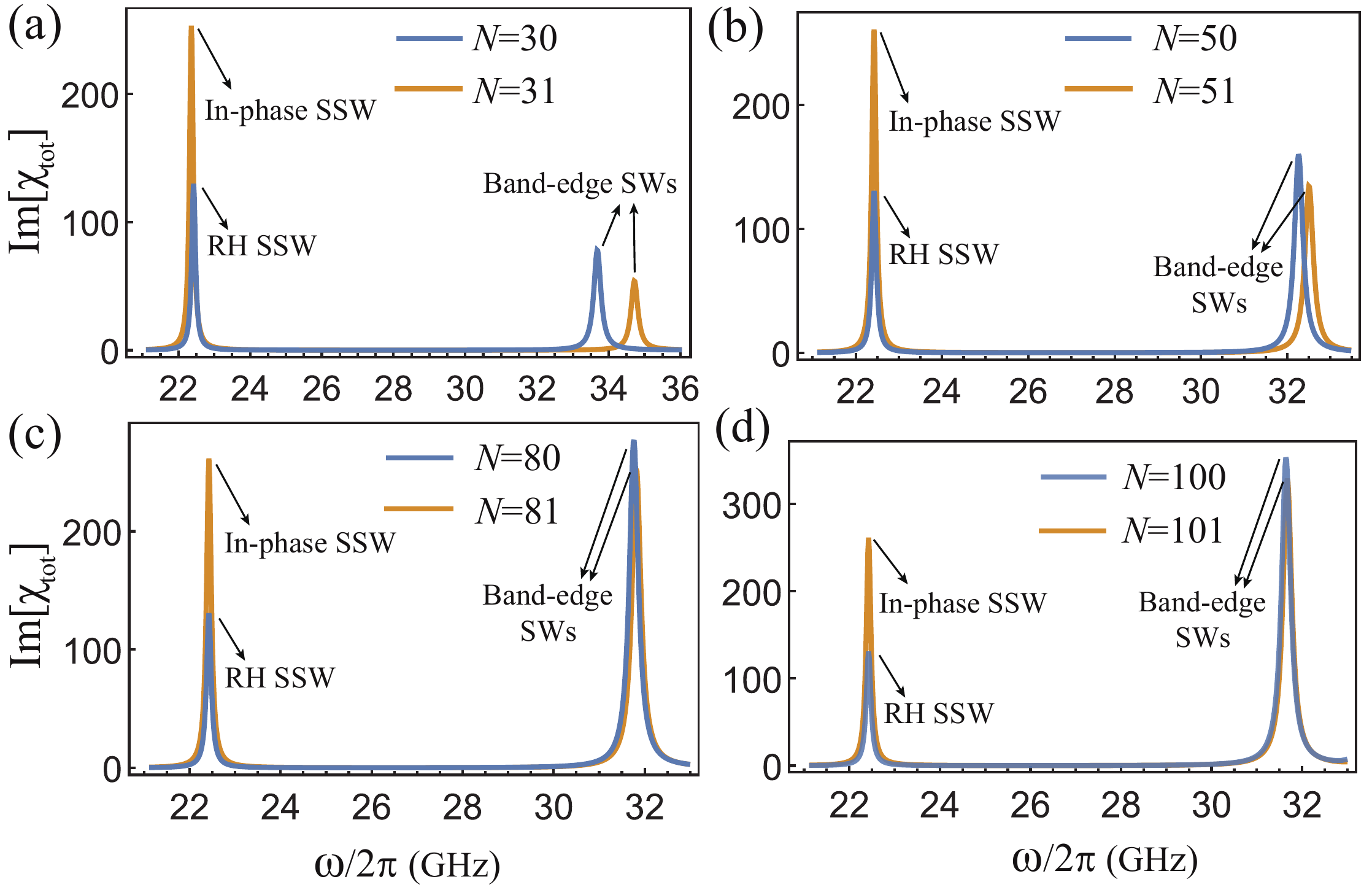}
	\caption{Absorption spectra of CrPS$_4$ with \textit{N} monolayers under a right-handed (RH) circularly polarized  microwave drive. We observe two peaks caused by surface spin waves (SSWs) and band-edge macrospin modes. In odd-numbered multilayers, only the in-phase SSW absorbs the radiation, whereas the out-of-phase one is inert owing to the lack of a net dynamic magnetization. Here the Gilbert damping $\alpha_{G}=10^{-3}$ \cite{freeman2025tunable}.  }
	\label{Fig2}	
\end{figure}

\textit{Microwave absorption.---}Right (left)-handed spin waves can be selectively excited by a right (left)-handed circularly polarized external \textit{rf} field. The resonant excitation of the surface modes manifests itself as a peak in the absorption spectrum within the magnon gap, at frequencies below that of the conventional (anti)ferromagnetic resonance. For uniform ($\mathbf{q}=0$) circularly polarized microwaves, $
\mathbf{h}_{\text{rf}}=h_{0}[\hat{\mathbf{x}}\cos\omega t+\hat{\mathbf{y}}\sin\omega t]$, the absorption power per unit area reads
\begin{equation}
P_{\text{abs}}=\frac{t_{0}}{2} \sum_{l}\mathbf{h}_{\text{rf}}\cdot\partial_{t}\mathbf{M}_{l}=\frac{1}{2}\omega h_{0}t_{0}M_{s}\sum_{l}\text{Im} [\widetilde{A}_{l}]. \label{Pab}
\end{equation}
Here $t_{0}$ is the interlayer distance and $\hat{\widetilde{A}}=(\widetilde{A}_{1},\widetilde{A}_{2},\cdots, \widetilde{A}_{N})^{T}$ the complex-valued spin-wave amplitude vector in the presence of damping and external driving, viz. 
\begin{equation}
\hat{\widetilde{A}}(\omega)=\frac{\gamma h_{0}}{\hat{\eta}\cdot(\hat{H}_{\mathbf{0}}-i\alpha_{G}\omega\hat{I})-\omega \hat{I}}\cdot  \hat{s} \label{A}
\end{equation}
where $\hat{s}=(1,-1,\cdots, (-1)^{N+1})^{T}$ encodes the interlayer antiferromagnetic configuration and $\hat{I}$ is the $N\times N$ identity matrix. Fig.~\ref{Fig2} plots the imaginary part of a dimensionless susceptibility $\chi_{\text{tot}}=(\mu_{0}M_{s}/h_{0})\sum_{l}\widetilde{A}_{l}$ as a function of frequency for the RH ($\omega>0$) \textit{rf} field, where $\mu_{0}$ is the vacuum permeability. 

The absorption power is governed by the number of in-phase interacting spins of the mode in question. In odd-numbered multilayers, only the in-phase surface mode can be excited, while the out-of-phase one is ``dark", i.e., it does not couple to the external \textit{rf} field due to the absence of a net oscillating magnetization. From Fig.~\ref{Fig2} the absorption of the in-phase surface mode exhibits a pronounced even-odd dependence on $N$: in odd-\textit{N} samples, the absorption is roughly twice that in even-\textit{N} samples, simply because the number of active spins is doubled for the in-phase surface mode of the uncompensated (odd-$N$) samples. In contrast, the absorption of the bulk band-edge mode becomes proportional to $N$ because of a larger number of interacting spins in thicker samples and the even-odd contrast is much weaker. Notably, the surface mode signals depend weakly on the specific value of odd- or even-\textit{N}, and exceed those of the band-edge mode up to $N\sim80$. 

\begin{figure}
	\includegraphics[width=7.2cm]{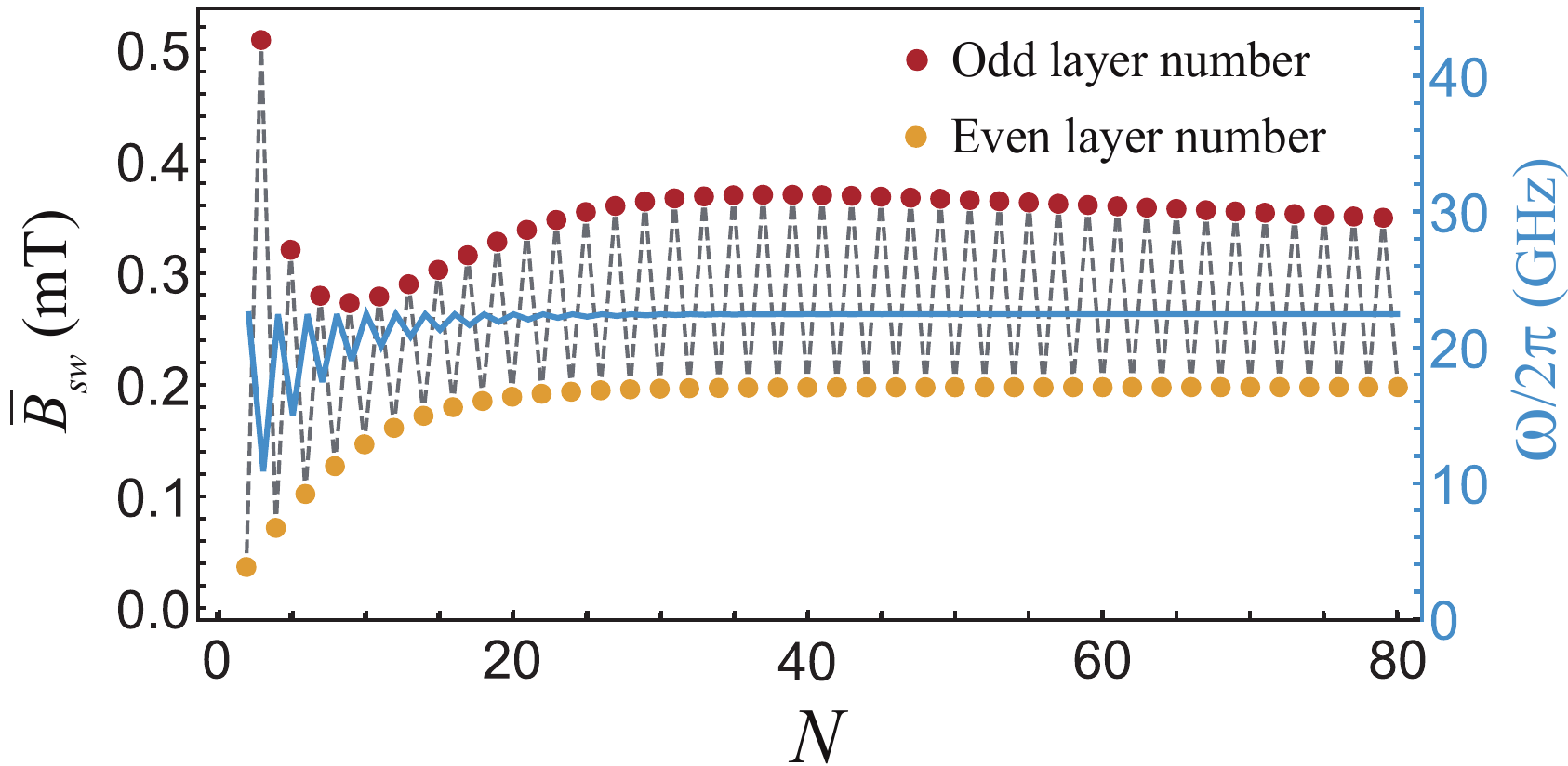}
	\caption{Layer number dependence of the stray field amplitude at a distance $d=30\,$nm above the top surface of CrPS$_4$ multilayers with interlayer spacing $t_{0}=6.11\,$\AA~\cite{PhysRevB.102.024408}, emitted by the lowest-frequency right-handed mode when resonantly excited by a right-handed microwave field with amplitude $h_{0}=1\,$mT and $q=6\times 10^{4}\,\text{cm}^{-1}$. The blue curve indicates the respective microwave drive frequencies.}
	\label{Fig3}	
\end{figure}

\textit{Magnetic stray fields.---}Excited spin waves radiate magnetic stray fields at their eigenfrequencies and with an amplitude that depends on their spatial distribution. Here, we predict an odd-even effect in the stray field of the surface mode, which should be observable by magnetometry of terraced vdW samples. According to Coulomb's Law, the stray field \(\mathbf{B}_{\text{sw}}\) at $\mathbf{r}=(\boldsymbol{\rho},z)$ emitted by a spin wave mode with amplitudes \(\mathbf{m}_l\) in layer \textit{l} reads~\cite{engel2005calculation,mahmoud2020introduction}
\begin{align}
\mathbf{B}_{\text{sw}}(\mathbf{r},t)=-\frac{\mu_{0}}{4\pi}t_{0} M_{s}\boldsymbol{\nabla} \displaystyle\sum_{l=1}^N\int d\boldsymbol{\rho}\frac{\mathbf{m}_l(\mathbf{r}_l,t)\cdot (\mathbf{r}-\mathbf{r}_l)}{|\mathbf{r}-\mathbf{r}_l|^3}.
\label{stray field1}
\end{align}
Assuming axial symmetry around the $z$ axis, we may choose $\mathbf{q}=q\hat{\mathbf{x}}$ without loss of generality. Substituting the spin-wave solution excited by the \textit{rf} field into Eq.~(\ref{stray field1}) leads to 
\begin{align}
\mathbf{B}_{\text{sw}}(x,z,t)=-\bar{B}_{sw}(q,z) \cos (qx-\omega t+\phi)\hat{\mathbf{x}} \label{Bsw}
\end{align}
where $\bar{B}_{sw}(q,z)=\frac{1}{2}\mu_{0} qt_{0}M_s \vert \widetilde{A}_{\text{tot}}\vert$ is the modulus and $\phi$ is the phase of the spin-wave amplitudes weighed according to their distance to the observation point:
\begin{equation}
 \widetilde{A}_{\text{tot}}(q,z)=\displaystyle\sum_{l=1}^{N}\widetilde{A}_{l}e^{-\vert q\vert\vert z-z_l\vert} .\label{Atot}
\end{equation}
The stray fields do not vanish for spin waves with finite wave vector $\mathbf{q}$ that can be excited, for instance, by focused laser beams as well as the \textit{rf} fields generated by narrow coplanar waveguides or striplines. For a spatially uniform, circularly polarized \textit{rf} field with constant amplitude $h_{0}$, $\widetilde{A}_{l}$ in Eq.~(\ref{Atot}) follows from Eq.~(\ref{A}) by performing the substitution $\hat{H}_{\mathbf{0}}\rightarrow \hat{H}_{\mathbf{q}}$. Fig.~\ref{Fig3} shows the $N$-dependence of the stray field ($\bar{B}_{sw}$) emitted by the lowest-frequency RH mode at a probing distance \(d = 30 \,\)nm above the topmost layer excited by a microwave field with $h_{0}=1\,$mT and $q=6\times 10^{4}\,\text{cm}^{-1}$. When $N\gtrsim30$, this mode becomes a well-localized surface mode with a stable frequency $\sim2\pi\times 22.5\,$GHz. The stray fields as a function of \textit{N} exhibit strong even-odd oscillations that are mainly caused by the oscillatory absorption power in compensated and uncompensated multilayers noted above. For even-numbered multilayers, $\bar{B}_{sw}$ gradually increases with $N$ and saturates at $N\gtrsim30$ indicating that the localized surface mode does not change anymore for thicker layers. In contrast, for odd-numbered multilayers, $\bar{B}_{sw}$ varies non-monotonically with the sample thickness ($N$): for $N < 30$, it first decreases and then increases with $N$. This effect is primarily caused by the layer thickness-dependent absorption of the driving microwaves. However, for odd-$N>31$, $\bar{B}_{sw}$ slightly decreases again with $N$ because the bottom surface contributes less due to the larger distance from the probe. In comparison, the even-odd variation in the stray field of bulk modes (not shown) becomes negligible for large $N$, since their spin-wave amplitudes are nearly uniformly distributed across layers. These features should offer a powerful tool to experimentally distinguish the proposed surface modes from bulk excitations.

\textit{Conclusions.---}We predict exchange-driven surface spin waves in vdW antiferromagnetic multilayers with ferromagnetic intralayer and antiferromagnetic interlayer coupling. In contrast to the magnetodipolar Damon-Eshbach waves, they remain robust in the short-wavelength regime and persist down to film thicknesses of a few tens of monolayers, and manifest as a strong low-frequency satellite in the microwave absorption spectrum. The magnetic stray fields emitted by the surface waves exhibit a pronounced even-odd oscillation with the number of layers that can be detected by NV-center magnetometry. While we focus here on coherent excitation by microwave fields, the surface modes can also be thermally excited and should be observable in thermodynamic properties, equilibrium and non-equilibrium noise spectra, and magnon transport experiments. Their distinct spatial and spectral properties may have important implications for the design and operation of ultrathin spintronic devices based on vdW magnetic materials. We expect that exchange interaction-based surface spin waves are generic to a broad class of magnetic films; however, the intrinsic atomic flatness of the vdW magnets studied here offers the advantage of minimal surface roughness (in each terrace)—an otherwise detrimental factor that could obscure or suppress these exchange surface modes.

\textit{Acknowledgment.---}We thank Toeno van der Sar and Yaroslav Blanter for valuable discussions and pointing us to CrPS$_4$. P.T. and G.B. acknowledge financial support by JSPS KAKENHI Grants Nos. 22H04965, 19H00645 and 24H02231. P. T. was also supported by JSPS KAKENHI Grant for Early-Career Scientists No. 23K13050, and Z.J.S. by the Chinese Scholarship Council under No. 202306130146.

\bibliography{reference}

\end{document}